\documentclass[runningheads]{llncs}
\usepackage[T1]{fontenc}
\usepackage{amsmath,amsfonts}
\usepackage{bbding}
\usepackage[numbers,sort&compress]{natbib}
\usepackage{hyperref}
\usepackage{graphicx}
\usepackage{subfigure}
\usepackage{xcolor}
\usepackage{multirow}
\newcommand{\xreal}{\boldsymbol{x}\textsubscript{real}}
\newcommand{\xfake}{\boldsymbol{x}\textsubscript{syn}}

\newcommand{\vz}{\boldsymbol{z}}
\newcommand{\vf}{\boldsymbol{f}}
\newcommand{\vw}{\boldsymbol{w}}

\begin{document}
%
\title{Deep Learning for Cancer Prognosis Prediction Using Portrait Photos by StyleGAN Embedding}
\titlerunning{Cancer Prognosis Prediction Using Portrait Photos by StyleGAN Embedding}
%
\author{
Amr Hagag\inst{1,2}
\and Ahmed Gomaa\inst{1,3}  \and Dominik Kornek\inst{1,3} \and Andreas Maier\inst{2} \and Rainer Fietkau\inst{1,3} \and Christoph Bert\inst{1,3} \and 
Yixing Huang\inst{1,3,}\Envelope \and Florian Putz\inst{1,3} }
\authorrunning{Hagag et al.}
%
\institute{
Department of Radiation Oncology, University Hospital Erlangen, Friedrich-Alexander-Universit\"at Erlangen-N\"urnberg, Erlangen, Germany\\
\email{yixing.yh.huang@fau.de}\\
 \and
Pattern Recognition Lab, Friedrich-Alexander-Universit\"at Erlangen-N\"urnberg, Erlangen, Germany\\
\and
Comprehensive Cancer Center Erlangen-EMN (CCC ER-EMN), Erlangen, Germany
}
\maketitle              
\begin{abstract}
Survival prediction for cancer patients is critical for optimal treatment selection and patient management. Current patient survival prediction methods typically extract survival information from patients' clinical record data or biological and imaging data. In practice, experienced clinicians can have a preliminary assessment of patients' health status based on patients' observable physical appearances, which are mainly facial features. However, such assessment is highly subjective. 
In this work, the efficacy of objectively capturing and using prognostic information contained in conventional portrait photographs using deep learning for survival prediction purposes is investigated for the first time.
A pre-trained StyleGAN2 model is fine-tuned on a custom dataset of our cancer patients' photos to empower its generator with generative ability suitable for patients' photos. 
The StyleGAN2 is then used to embed the photographs to its highly expressive latent space.
Utilizing state-of-the-art survival analysis models and StyleGAN’s latent space embeddings, this approach predicts the overall survival for single as well as pan-cancer, achieving a C-index of $0.680$ in a pan-cancer analysis, showcasing the prognostic value embedded in simple 2D facial images.
In addition, thanks to StyleGAN's interpretable latent space, our survival prediction model can be validated for relying on essential facial features, eliminating any biases from extraneous information like clothing or background. 
Moreover, our approach provides a novel health attribute obtained from StyleGAN’s extracted features, allowing the modification of face photographs to either a healthier or more severe illness appearance, which has significant prognostic value for patient care and societal perception, underscoring its potential important clinical value.

\keywords{Survival prediction \and cancer \and StyleGAN \and deep learning \and latent space \and face prognosis \and explainable AI.}
\end{abstract}
\section{Introduction}
Survival analysis for cancer patients is critical for optimal treatment selection and patient management. Patients with very adverse prognosis may need aggressive management and acute intervention. 
However, patients with good prognostic constellations similarly may benefit from specific treatment approaches like dose escalation and radical treatment of all sites in radiotherapy. So far, a vast class of survival prediction models have been investigated, ranging from conventional regression algorithms \cite{emura2012survival,kidd2018survival} to the latest radiomics \cite{van2017survival,chen2022mri} and deep learning methods \cite{wang2021cox,katzman2018deepsurv,kim2019deep,wankhede2022dynamic}.
With the successful applications in various fields, nowadays deep learning has become the main trend for patient survival prediction, because of its high representation power and ability to extract essential features from various types of data. 
Nevertheless, all the previous methods rely on data ranging from regular clinical records to commonly used medical imaging data (e.g., MRI and CT) as well as biological data (e.g., histologic and genetic information) \cite{vale2021long}. 
Within radiotherapy practice, the acquisition of patient photographs is a routine measure implemented for identity verification purposes.
Nevertheless, the potential utility in predicting patient survival and overall health status based on facial features derived from such widely available 2D portrait photos remains unexplored.

Facial features observable in 2D portrait photographs, like skin colorit and facial expression can reflect underlying pathophysiologic processes like muscle wasting, lowering of muscle tone or changes in microcirculation \cite{oken1982toxicity}. These features therefore hold considerable potential as indicators of a patient's underlying health status.
Many severe diseases will lead to evident facial changes, e.g., the most well-known diseases being genetic ones like the Down's syndrome \cite{gurovich2019identifying}. 
Clinicians with extensive experience frequently utilize patients' physical appearance, as well as other related features, to estimate survival time and evaluate overall functional status. The Eastern Cooperative Oncology Group (ECOG) score is such a widely used metric for evaluating patient health status \cite{oken1982toxicity}. 
Hui et al. \cite{hui2015diagnostic} were able to show that drooping of the nasolabial fold  was a highly predictive indicator of imminent death within a three-day timeframe.
Patient appearance has also been shown to be a factor in determining healthcare priority in emergency situations \cite{bagnis2020judging}. However, assessments made by clinicians are inherently subjective and can therefore exhibit a high degree of intra- and inter-rater variance. In this work, we aim to achieve an objective assessment of cancer patients' prognosis based on facial features using deep learning.
 
Deep learning-based facial image analysis can reveal patient health status caused by various factors \cite{su2021deep}. It has been reported to be effective in detecting neurological disorders \cite{yolcu2017deep}, acromegaly \cite{kong2018automatic}, genetic disorders \cite{gurovich2019identifying}, and cancer \cite{liang2020identification}.
  However, deep learning models can be biased \cite{wallis2022clever} and have poor generalizability. 
  To ensure meaningful facial features are used, interpretability of the model is crucial. StyleGAN \cite{karras2020training,karras2020analyzing} is a widely used network for face image synthesis due to its visual fidelity, diversity, and high editability over its latent space. 
  Essential prognostic features can potentially be embedded into the latent space as well and hence it presents a promising avenue for survival analysis. A notable advantage of leveraging the latent space for prognosis prediction is the ability to visually and qualitatively evaluate the associations between the features within the latent space and patient's prognosis.
Our contributions can be summarized as follows: a) To the best of our knowledge, this is \textbf{the first attempt} to use deep learning for cancer patient survival prediction based on 2D facial photographs;
b) We propose a model that utilizes the StyleGAN expressive latent space feature representations for survival prediction.
c) We introduce a \textbf{novel addition} to traditional prognostic methods by using StyleGAN embeddings alongside clinical parameters to improve survival prediction accuracy.
d) To provide \textbf{interpretability and validation} to the model, a ``health" attribute is extracted to allow adjusting face photographs to either a healthier or more severe illness appearance. This manipulation has significant prognostic value for patient care and societal perception, underscoring its potential important clinical value.


\begin{figure}[b]
\begin{minipage}[b]{0.16\linewidth}
\subfigure[$\xreal$]{
\includegraphics[width=\linewidth]{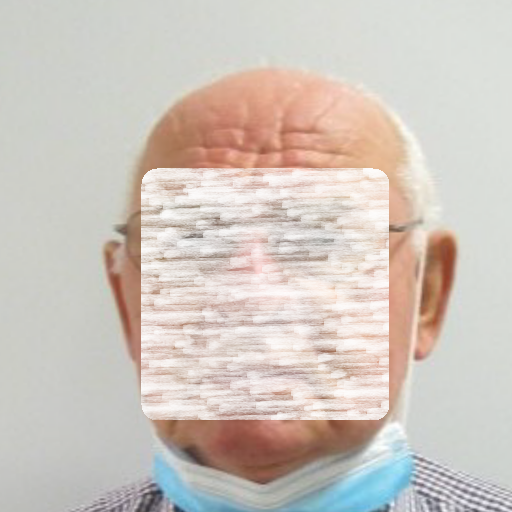}
}
\end{minipage}
\begin{minipage}[b]{0.16\linewidth}
\subfigure[$\xfake$ \tiny{w/o FT}]{
\includegraphics[width=\linewidth]{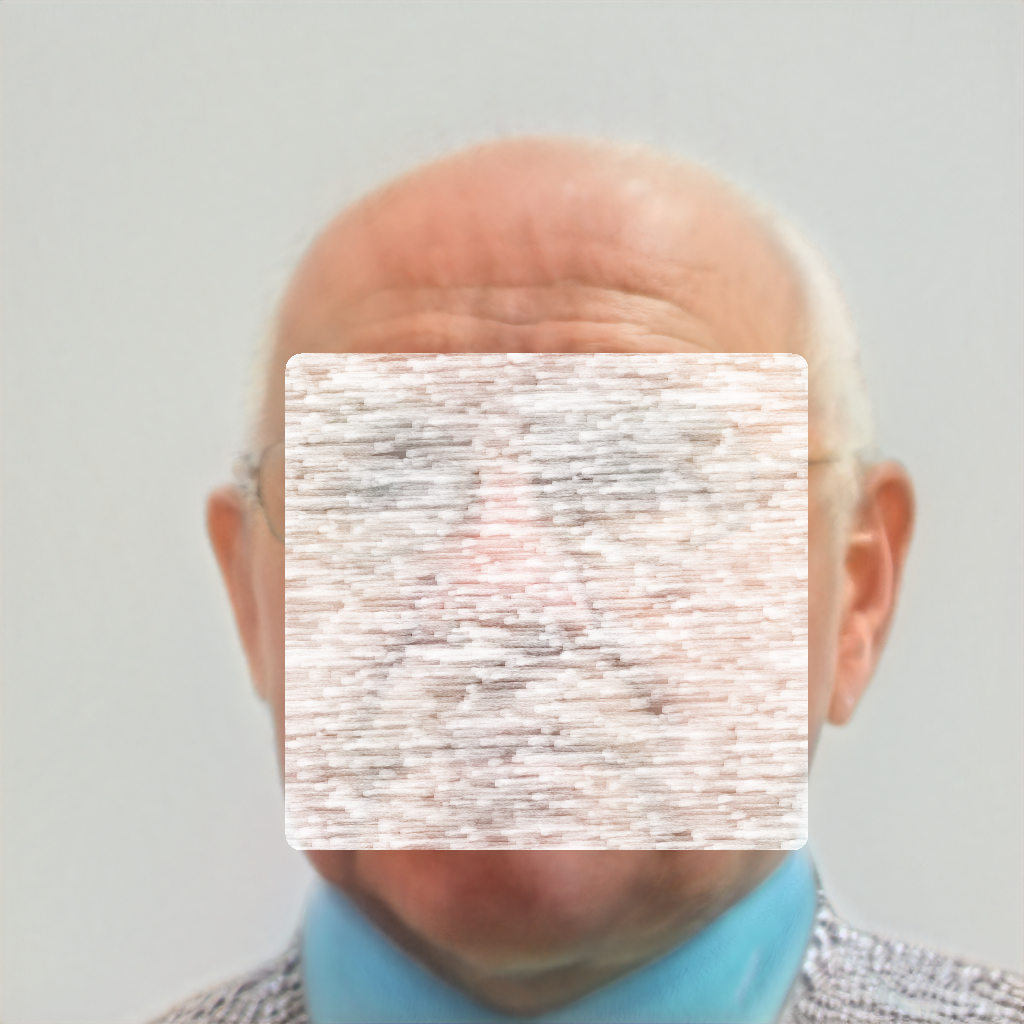}
\label{subfig:noFT1}
}
\end{minipage}
\begin{minipage}[b]{0.16\linewidth}
\subfigure[$\xfake$ \tiny{with FT}]{
\includegraphics[width=\linewidth]{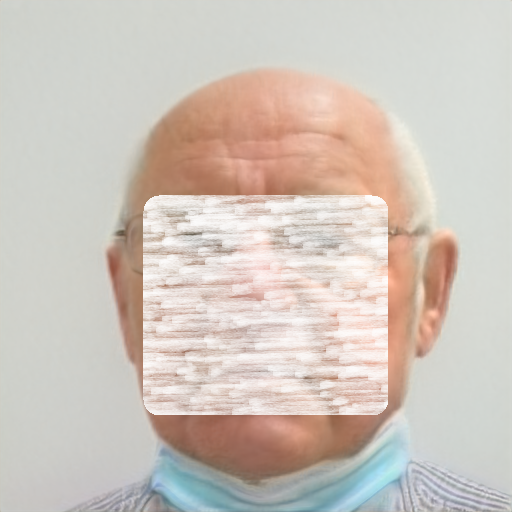}
\label{subfig:FT1}
}
\end{minipage}
\begin{minipage}[b]{0.16\linewidth}
\subfigure[$\xreal$ ]{
\includegraphics[width=\linewidth]{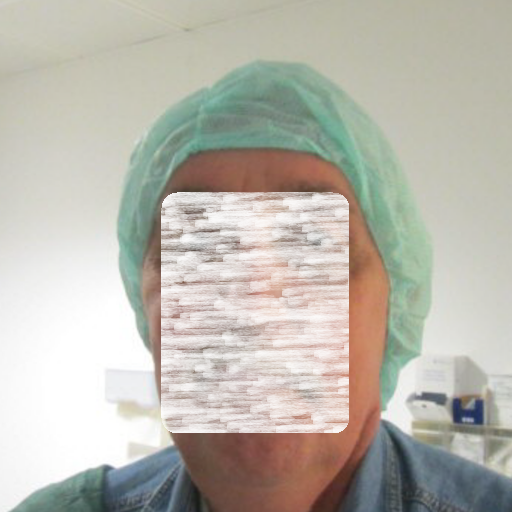}
}
\end{minipage}
\begin{minipage}[b]{0.16\linewidth}
\subfigure[$\xfake$ \tiny{w/o FT}]{
\includegraphics[width=\linewidth]{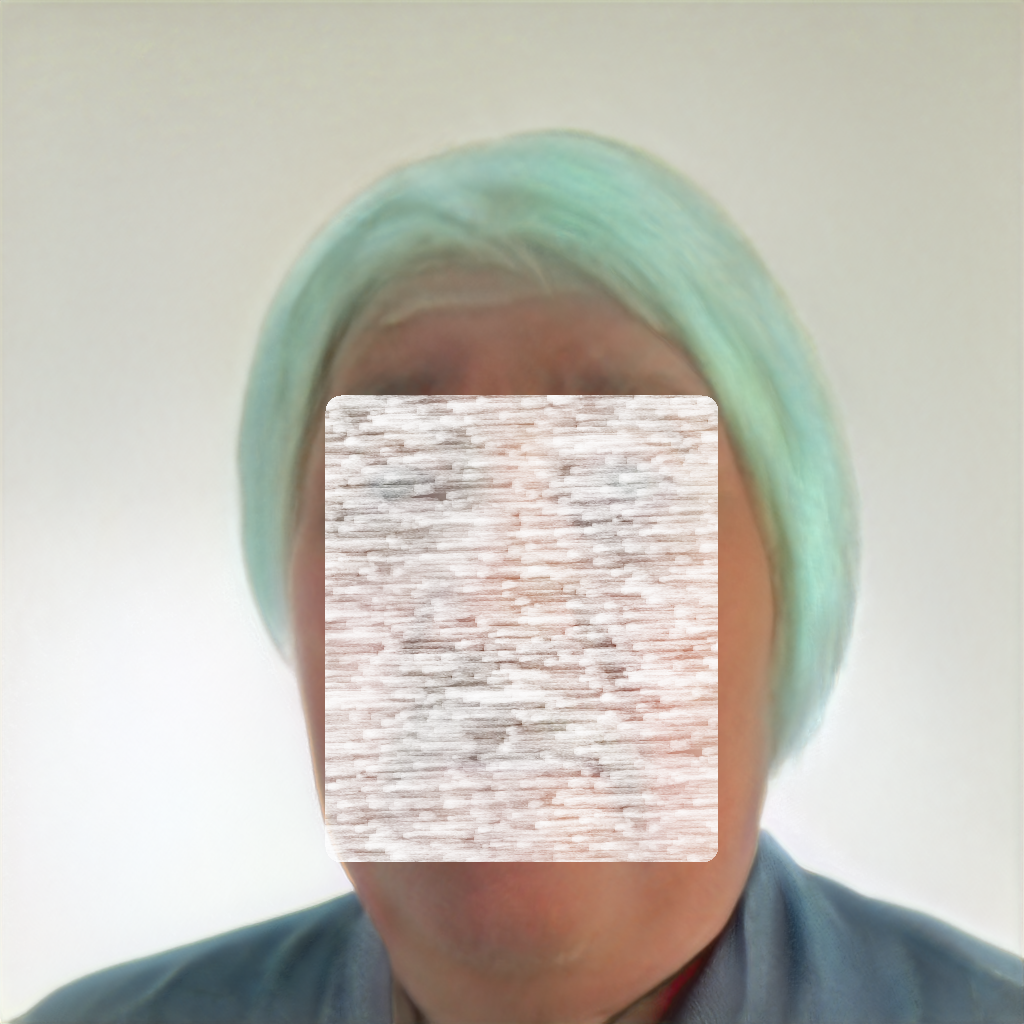}
\label{subfig:noFT2}
}
\end{minipage}
\begin{minipage}[b]{0.16\linewidth}
\subfigure[$\xfake$ \tiny{with FT}]{
\includegraphics[width=\linewidth]{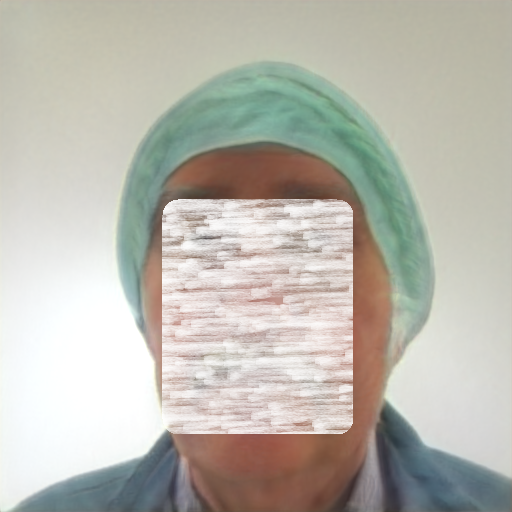}
\label{subfig:FT2}
}
\end{minipage}
\caption{Exemplary anonymized face images and their corresponding synthetic images generated by StyleGAN without (w/o) and with fine-tuning (FT).}
\label{Fig:FineTuningEffect}
\end{figure}

\section{Material and Methods}
\textbf{2.1 Data}
\vspace{5pt}

The dataset utilized in this work comprises 13,503 2D portrait photographs captured during the admission of individuals who have been diagnosed with cancer. Of this cohort, 52.5\% are male and 47.5\% are female. Notably, 57.3\% of the dataset is right censored. In addition, the survival times for the non-censored patients range from 2 to 4,791 days. To mitigate potential biases in the photographs due to factors such as clothing and background, an automated face recognition process utilizing dlib \cite{dlib09} is employed to align and crop all images to a standardized size of $512 \times 512$. The clinical information about the patients is retrieved, including international classification of diseases (ICD), gender, age, number of tumors, etc. Information acquired after the photo acquisitions is excluded. 

\noindent\textbf{Data Anonymization Disclaimer:} This work's data is anonymized to safeguard individual privacy. Facial images are synthetic, anonymized, or expert-reviewed projections, ensuring individuals cannot be identified. 


\vspace{5pt}
\noindent\textbf{2.2 Embedding Images to StyleGAN2 Latent Space}
\vspace{5pt}

To accurately capture patients' facial features from a given photo, a generative model should be able to accurately reproduce these expressive features into a generated image.
A StyleGAN2 model pretrained on the FFHQ dataset \cite{karras2019style} is fine-tuned using adaptive discriminator augmentation (ADA) \cite{karras2020training} with our custom dataset of cancer patients' face images. 
This step is necessary to increase StyleGAN's generation ability for the special facial features in cancer patients and auxiliary medical equipment such as masks, breathing tubes and gauze. Fig.\,\ref{Fig:FineTuningEffect} shows the difference between synthesised images with and without fine-tuning. Without fine-tuning, the pretrained model fails to generate the mask and the bouffant cap in Fig.\,\ref{subfig:noFT1}  and Fig.\,\ref{subfig:noFT2}, respectively.


\noindent\textbf{Photo Embedding:} To project images from the image space to
the latent space, for each photo $\xreal$, an optimization process searches via gradient decent for a $512$-dimensional latent space vector $\vz$ if input to the StyleGAN2 generator, will produce a corresponding synthetic image $\xfake$ similar to the original image $\xreal$. This similarity is defined as Euclidean distance in the feature space of a pretrained VGG-16,
\begin{equation}
\vz = \arg \min_{\vz} \phi\left(\xreal, \xfake \right) = \arg \min_{\vz} \phi\left(\xreal, G(\vz) \right) ,
\end{equation}
where $G$ is the generator of the fine-tuned StyleGAN2 model, and $G(\vz)$ is the synthetic image $\xfake$, and $\phi$ is the loss function. 
In such a way, each face image is associated with one optimized highly expressive $512$-dimensional latent vector.

\vspace{5pt}
\noindent\textbf{2.3 Survival Analysis Models}
\vspace{5pt}

Survival analysis models are used to predict patients' survival by estimating the hazard function, which is the instantaneous rate at which the event of interest (e.g., death) occurs at a given time, and allows for the inclusion of multiple covariates to predict the risk of the event. 
The most widely used survial model in practice is the Cox Proportional Hazard (CoxPH) \cite{lin2002modeling} model based on conventional clinical features.
In the CoxPH model, the hazard function has the following form \cite{katzman2018deepsurv},
\begin{equation}
\lambda(t|\vf)=\lambda_0(t)\cdot e^{h(\vf)}=\lambda_0(t)\cdot e^{\vw^\top\vf},
\end{equation}
where $\lambda_0(t)$ is a baseline hazard function, $\vf$ is the input feature vector, $h(\vf)$ is the log-risk function, and $\vw$ is the coefficient vector to optimize. 

To learn a more complex, non-linear relationship between the features for the hazard estimation, deep learning networks have been proposed, with DeepSurv \cite{zhu2016deep,katzman2018deepsurv} being the state-of-the-art. The DeepSurv method applies a multi-layer perceptron (MLP) to estimate the log-risk function $h(\vf)$ of the CoxPH model. To extract features from medical images, the MLP network can be replaced by a convolutional neural network (CNN) \cite{zhu2016deep,gurovich2019identifying}. 
In this work, a ResNet-18 and a VGG-16 network are investigated as examples of the CNN with image data. 
As CNN-extracted features are highly abstract, in this work, we propose to integrate the StyleGAN network to the DeepSurv framework for the prediction of continuous survival time, i.e., utilizing the expressive StyleGAN photo embedding $\vz$ as the input features of the CoxPH and DeepSurv models.

\begin{figure}[t]
\includegraphics[width=1\linewidth]{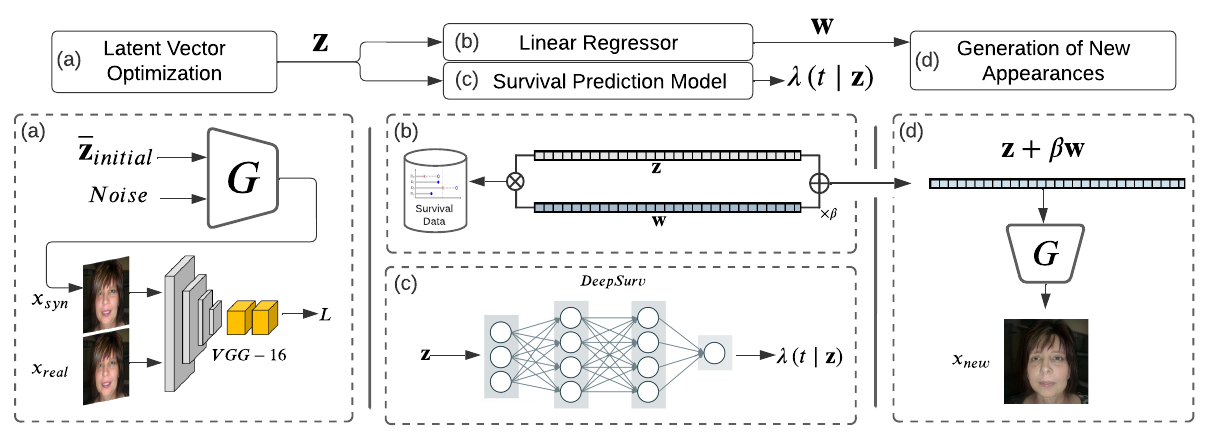}
\caption{The workflow of our proposed facial prognosis method: (a) Optimization of $\vz$ using StyleGAN generator fine-tuned on custom cancer patient data; (b) CoxPH regression to get $\vw$ for health attribute construction and latent space manipulation; (c) DeepSurv for survival prediction; (d) Generation of new appearance.}
\end{figure}

\vspace{5pt}
\noindent\textbf{2.4 Latent Space Manipulation}
\vspace{5pt}

In this work, we propose two tests to visually and qualitatively evaluate the semantic meaning of our photo embedding as well as the prognostic value it holds for the task of survival prediction. Each of the tests can be conducted by simple manipulations of the latent vector $\vz$. 

\noindent\textbf{Health Attribute Manipulation:} Leveraging the linearity of the CoxPH regression model, we fit the $512$-dimensional vectors $\vz$ to the known survival data,
\begin{equation}
h(\vz) =\sum_{i=0}^{N-1}\vw_i \vz_i,
\end{equation}
where $\vw$ is the vector of linear coefficients, and $h(\vz)$ is the log-risk function which is fed to the CoxPH model. By manipulating $\vz$ in all the dimensions jointly with a multiplier of $\vw$,
\begin{equation}
\hat{\vz} = \vz + \beta \vw,
\label{eqn:jointFeature}
\end{equation}
the dimensions in $\vz$ with high importance are modified to a large degree, while those with low importance stay almost unchanged. By changing $\beta$, the StyleGAN generator is able to generate new appearances of this patient from $\hat{\vz}$, with a healthier or more severe illness appearance. This vector $\vw$ serves as a health attribute.

\noindent\textbf{Age Manipulation:} To further validate the quality of our facial features vectors and ensure that our predictive model is focusing on complex survival related aspects and not only a single aspect like the age, we use a simple linear regression to map the feature vectors $\vz$ to the known age of the patients. Similar to (Eqn.\,(\ref{eqn:jointFeature})), $\vw$ here would serve as a simple age attribute.

\section{Results and Discussion}
\textbf{3.1 Pan-Cancer Analysis for Survival Prediction Results}

The survival prediction results are displayed in Table\,\ref{Tab:accuracy}. 
Our approach for continuous survival time prediction using DeepSurv based on StyleGAN photo embeddings achieved for pan-cancer analysis a mean concordance index (C-index) of $0.680$ and a mean Brier Score of $0.200$. 
To put this into perspective, the model's survival predictions are evidently higher than chance. 
This reflects that the model uses key survival features from relevant prognostic facial attributes embedded in the latent vectors. 
As also shown in Table\,\ref{Tab:accuracy}, the DeepSurv model using patients' portraits fed directly through a CNN achieved a C-index of $0.51$ and $0.47$ using the ResNet-18 and the VGG-16 networks respectively.
In this case, the CNNs failed to extract the needed prognostic information directly from the portrait photos. It is worth noting that the comparison methods (i.e., DS+ResNet and DS+VGG with input photo data) used end-to-end training instead of two-step training as our proposed method (i.e., DS+MLP with input combined embedding and clinical data).

As also shown in Table\,\ref{Tab:accuracy}, survival prediction models based on conventional clinical features achieved a C-index of 0.690 and 0.729 using CoxPH and DeepSurv models respectively, which is comparable to what was achieved by the embeddings of simple 2D portraits. 
Combining clinical features with photo embeddings using late fusion boosted the C-index to 0.787 highlighting again the complementary prognostic value of StyleGAN's embeddings. 


\begin{table}[t]
\caption{Survival prediction results of different methods using DeepSurv (DS) and CoxPH models. C-index and Brier score (B-Score) are calculated for all models.}
\label{Tab:accuracy}
\centering
\begin{tabular}{l|l r r||l| l r r}
\hline
\textbf{Input data}& \textbf{Method} & \textbf{C-index} & \textbf{B-Score} & \textbf{Input data} & \textbf{Method} & \textbf{C-index} & \textbf{B-Score} \\
\hline
Embeddings & DS+MLP &0.680 & 0.200 & Clinic. Data & CoxPH & 0.690 & 0.151\\ 
 
 Clinic. Data&DS+MLP  & 0.729 & 0.168 &Photos& DS+ResNet& 0.510 & 0.471\\
Combined& DS+MLP& 0.787 & 0.137 &Photos &DS+VGG & 0.470 & 0.471 \\
\hline
\end{tabular}
\end{table}

\begin{figure}[h]
\includegraphics[width=1\linewidth]{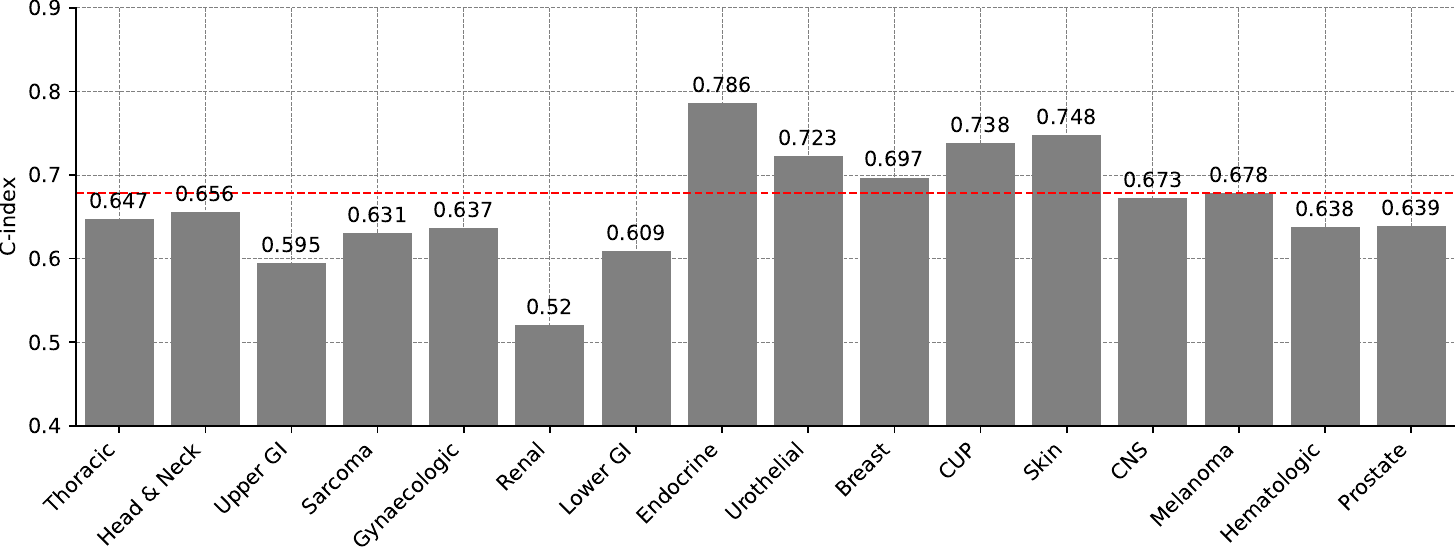}
\caption{A subgroup analysis on 16 cancer types with corresponding single C-indices. The dashed line represents the average C-index of the pan-cancer analysis.}
\label{Fig:subgroups}
\end{figure}

\noindent\textbf{3.2 Subgroup Analysis for Survival Prediction Results}

Intriguingly, as shown in Fig \,\ref{Fig:subgroups}, the developed pan-cancer model showed good prognostic accuracy for nearly all cancer sub-types, suggesting that prognostically-relevant facial features are present in most malignancies. 
However, there were also notable differences in observed prognostic accuracy. High prognostic accuracy was observed for endocrine malignancies (including small-cell cancers), metastatic cancer of unknown primary (CUP), urothelial cancer, breast cancer and melanoma, with C-indices of 0.786, 0.738, 0.723, 0.697 and 0.678, respectively, whereas notably lower accuracy was observed for lower and upper gastrointestinal cancers, prostate cancer, hematologic malignancies, sarcoma and especially renal cancer, with C-indices of 0.609, 0.595, 0.639, 0.638, 0.631 and 0.52, respectively. 
From a clinical perspective, this pattern could be derived from differences in tumor biology, as hematologic malignancies as well as sarcomas have a different origin and a distinct biology from carcinomas \cite{berman2005modern}. 
In addition, prostate cancer, colorectal cancer, and renal cancer are slowly proliferating tumors with markedly slower disease progression, lower tendency for metastasis and good systemic treatment options that typically result in patients showing less severe systemic signs of the disease \cite{capitanio2016renal}.

\begin{figure}[t]
\centering
\begin{minipage}[b]{0.16\linewidth}
\centering
$\beta = -10$
\end{minipage}
\begin{minipage}[b]{0.16\linewidth}
\centering
$\beta = -5$
\end{minipage}
\colorbox{lightgray}{
\begin{minipage}[b]{0.14\linewidth}
\centering
$\xfake, \beta = 0$
\end{minipage}}%
\begin{minipage}[b]{0.16\linewidth}
\centering
$\beta = 5$
\end{minipage}
\begin{minipage}[b]{0.16\linewidth}
\centering
$\beta = 10$ 
\end{minipage}
\begin{minipage}[b]{0.16\linewidth}
\centering
$\beta = 20$ 
\end{minipage}

\begin{minipage}[b]{0.166\linewidth}
\includegraphics[width=\linewidth]{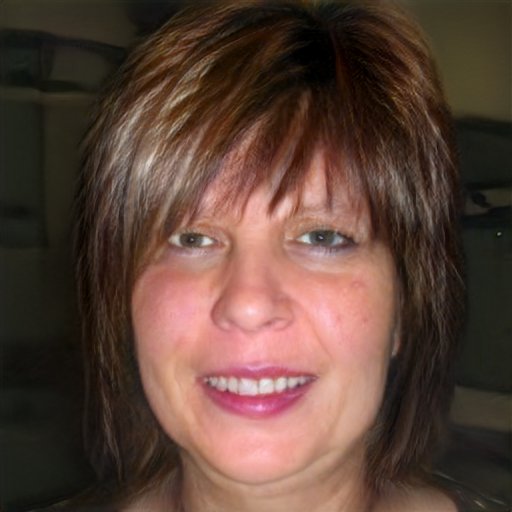}
\end{minipage}%
\begin{minipage}[b]{0.166\linewidth}
\includegraphics[width=\linewidth]{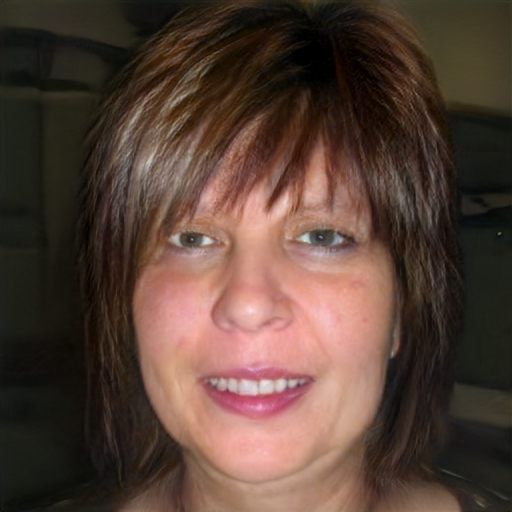}
\end{minipage}%
\begin{minipage}[b]{0.166\linewidth}
\includegraphics[width=\linewidth]{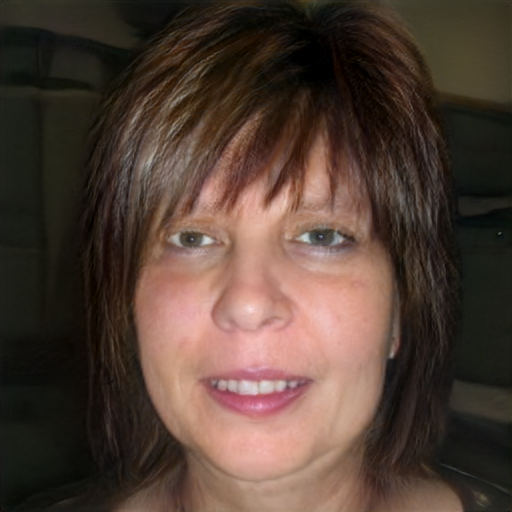}
\end{minipage}%
\begin{minipage}[b]{0.166\linewidth}
\includegraphics[width=\linewidth]{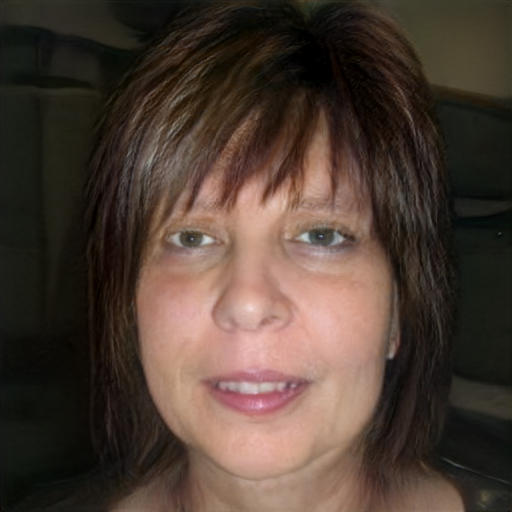}
\end{minipage}%
\begin{minipage}[b]{0.166\linewidth}
\includegraphics[width=\linewidth]{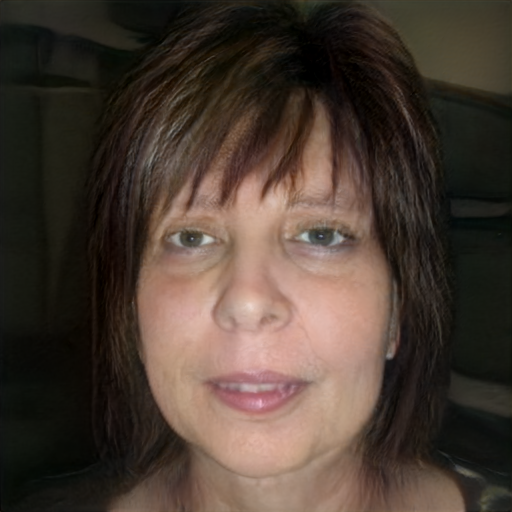}
\end{minipage}%
\begin{minipage}[b]{0.166\linewidth}
\includegraphics[width=\linewidth]{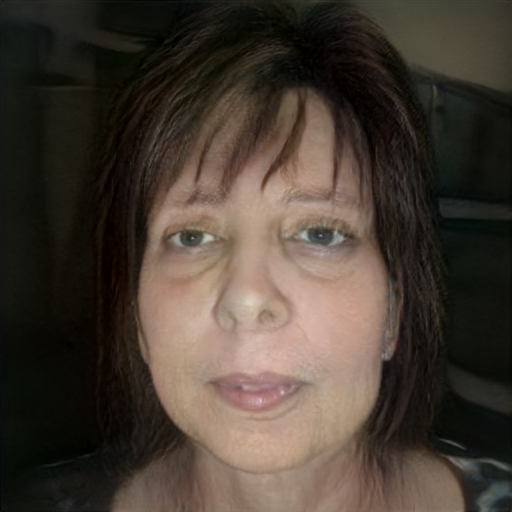}
\end{minipage}%

\begin{minipage}[b]{0.166\linewidth}
\includegraphics[width=\linewidth]{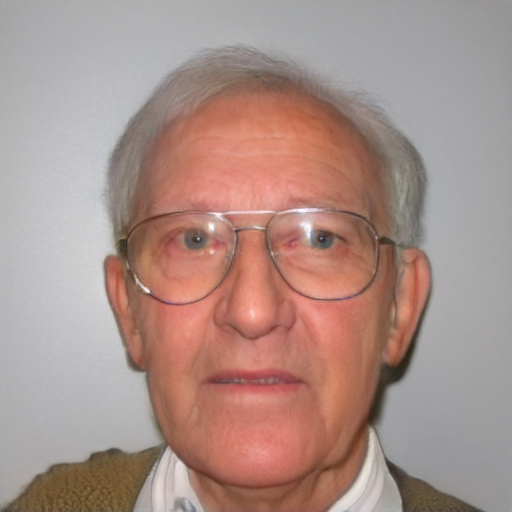}
\end{minipage}%
\begin{minipage}[b]{0.166\linewidth}
\includegraphics[width=\linewidth]{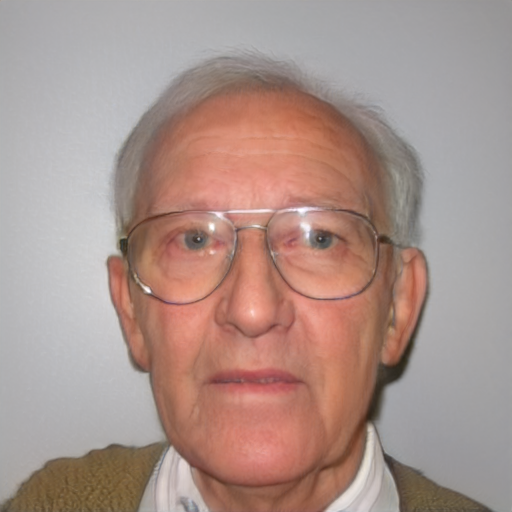}
\end{minipage}%
\begin{minipage}[b]{0.166\linewidth}
\includegraphics[width=\linewidth]{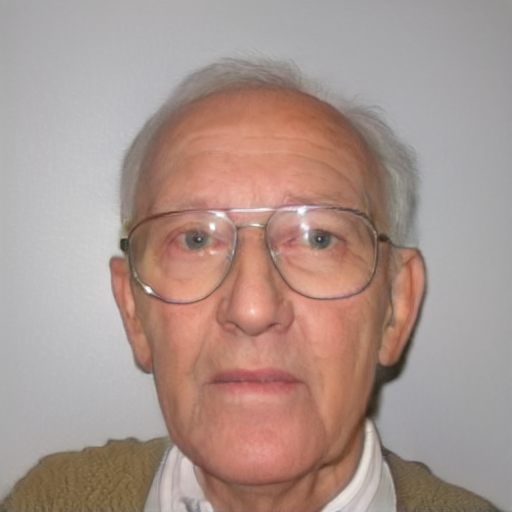}
\end{minipage}%
\begin{minipage}[b]{0.166\linewidth}
\includegraphics[width=\linewidth]{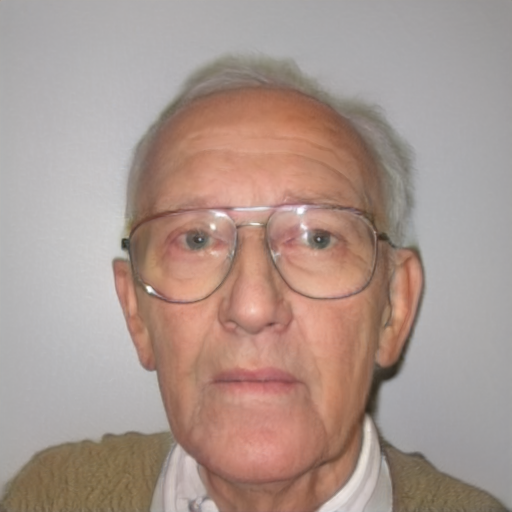}
\end{minipage}%
\begin{minipage}[b]{0.166\linewidth}
\includegraphics[width=\linewidth]{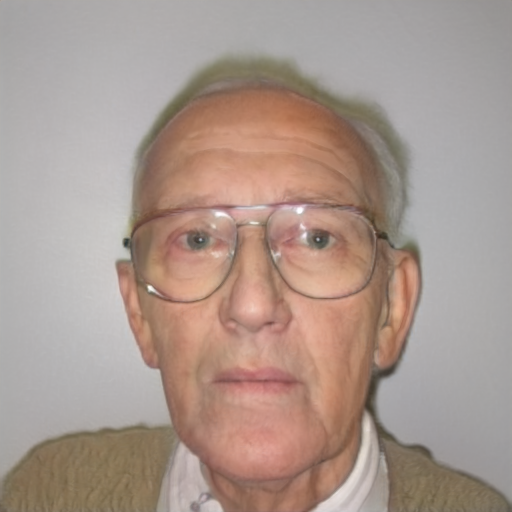}
\end{minipage}%
\begin{minipage}[b]{0.166\linewidth}
\includegraphics[width=\linewidth]{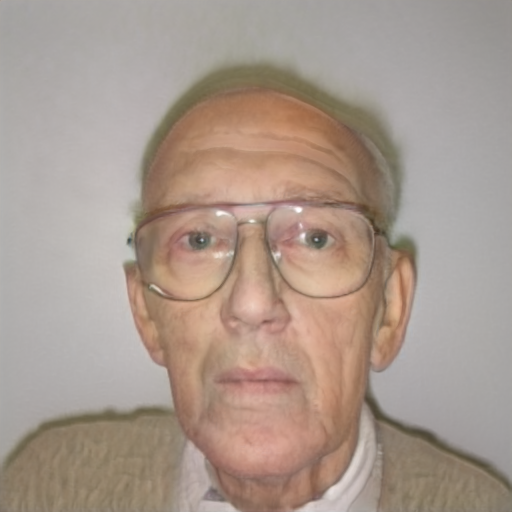}
\end{minipage}%

\caption{Manipulating the health attribute with different $\beta$ values for regenerated patients' images. The synthesized image without manipulation is at $\beta=0$ (third column).}
\label{Fig:JointFeatures}
\end{figure}
\vspace{5pt}
\noindent\textbf{3.3 Model Interpretability with Health Attribute}
\vspace{5pt}

Fig.\,\ref{Fig:JointFeatures} demonstrates the synergetic effect of manipulating patient's images in the latent space using the health attribute
 (Eqn.\,(\ref{eqn:jointFeature})). In Fig.\,\ref{Fig:JointFeatures} the appearances of the patients are controlled by $\beta$. With a negative value for $\beta$, the patients appear to be healthier with a reddish skin colorit, full cheeks, a smiling expression and a strong tone of the facial muscles.
 In contrast, with a positive value for $\beta$, there is a loss of masticatory muscles and subcutaneous fat with a pale skin colorit. Moreover, the tone of the facial muscles is reduced with drooping of the nasolabial fold, an overall more relaxed appearance of facial features and an increased ``tippiness" of the nose. These features are consistent with long-standing clinical descriptions of prognostically relevant changes in facial appearance \cite{hui2015diagnostic, withington2005hippocratic}. Most importantly, the change is mainly located in the head region instead of their clothes or the background. The model is also able to reflect patients' mental status to some degree by extracting their facial expressions.
 
 This editability of patients' photos to have either a healthier appearance or prognosis of a more severe illness has important value for patient care. For cancer patients, the quality of life is as important as the prevention, treatment and management of their illness. The quality of life relies on not only physical but also mental well-being \cite{rankin2003perceived,de2017quality}. Editing their photos to have a healthier appearance and sharing such optimized photos to their social networks can convey a more positive and optimistic message. For healthcare professionals, this could enable illustrating a patients' appearance in a healthy state or a state of illness progression, which could improve empathy and caretaking.
 
\begin{figure}[t]
\centering
\begin{minipage}[b]{0.16\linewidth}
\centering
$\beta = -10$
\end{minipage}
\begin{minipage}[b]{0.16\linewidth}
\centering
$\beta = -5$
\end{minipage}
\fboxsep0pt
\colorbox{lightgray}{
\begin{minipage}[b]{0.15\linewidth}
\centering
$\xfake, \beta = 0$
\end{minipage}}%
\begin{minipage}[b]{0.16\linewidth}
\centering
$\beta = 5$
\end{minipage}
\begin{minipage}[b]{0.16\linewidth}
\centering
$\beta = 10$ 
\end{minipage}
\begin{minipage}[b]{0.16\linewidth}
\centering
$\beta = 20$ 
\end{minipage}

\begin{minipage}[b]{0.166\linewidth}
\includegraphics[width=\linewidth]{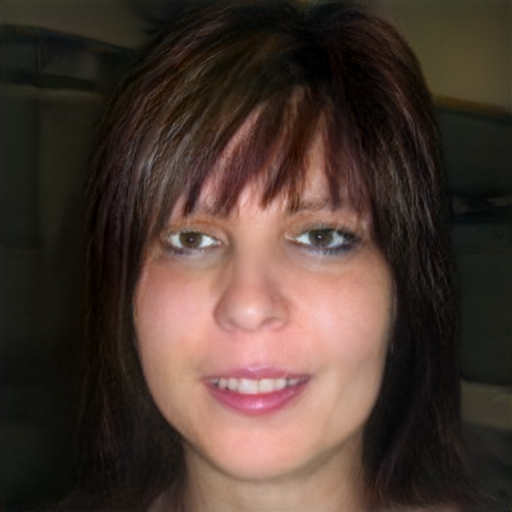}
\end{minipage}%
\begin{minipage}[b]{0.166\linewidth}
\includegraphics[width=\linewidth]{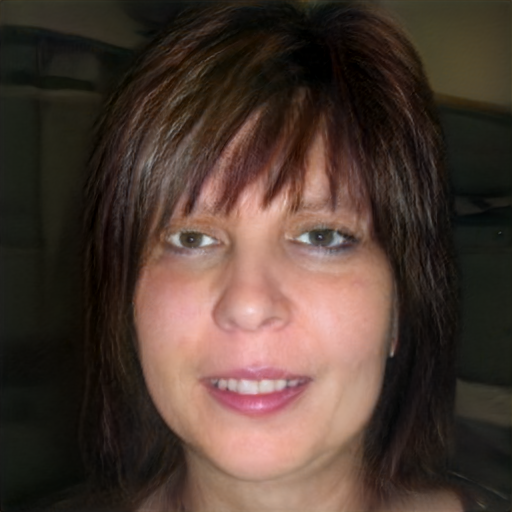}
\end{minipage}%
\begin{minipage}[b]{0.166\linewidth}
\includegraphics[width=\linewidth]{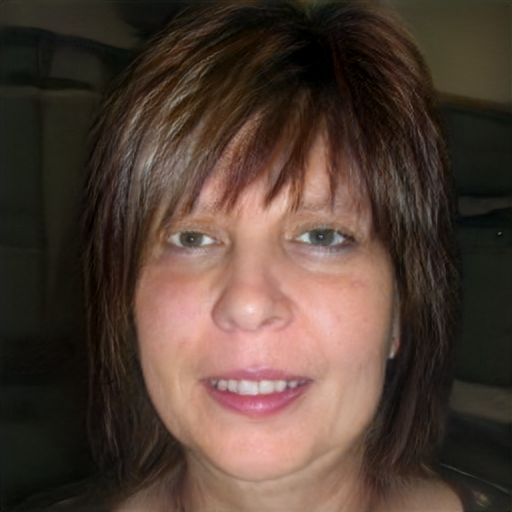}
\end{minipage}%
\begin{minipage}[b]{0.166\linewidth}
\includegraphics[width=\linewidth]{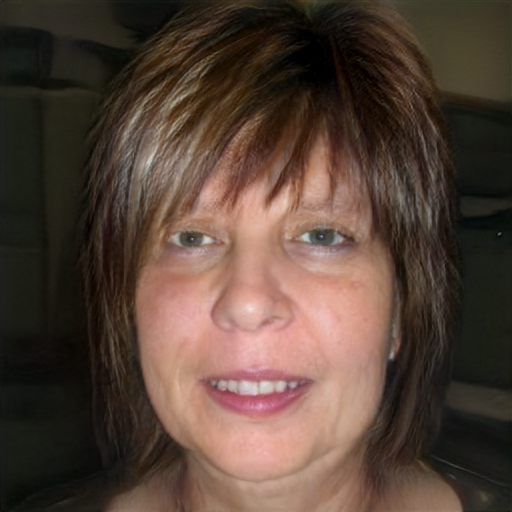}
\end{minipage}%
\begin{minipage}[b]{0.166\linewidth}
\includegraphics[width=\linewidth]{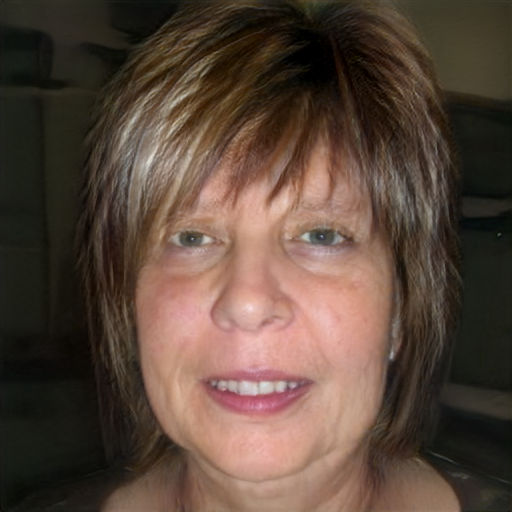}
\end{minipage}%
\begin{minipage}[b]{0.166\linewidth}
\includegraphics[width=\linewidth]{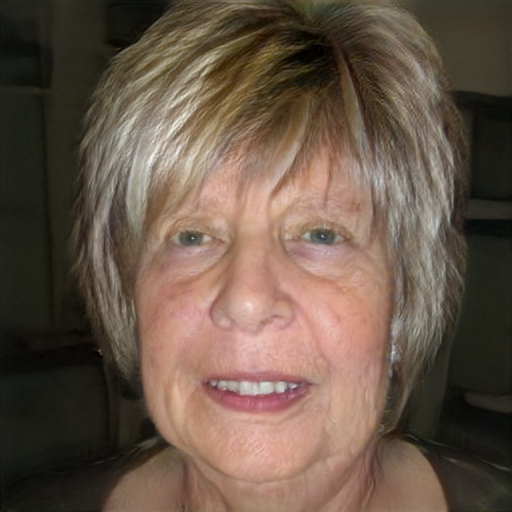}
\end{minipage}%

\begin{minipage}[b]{0.166\linewidth}
\includegraphics[width=\linewidth]{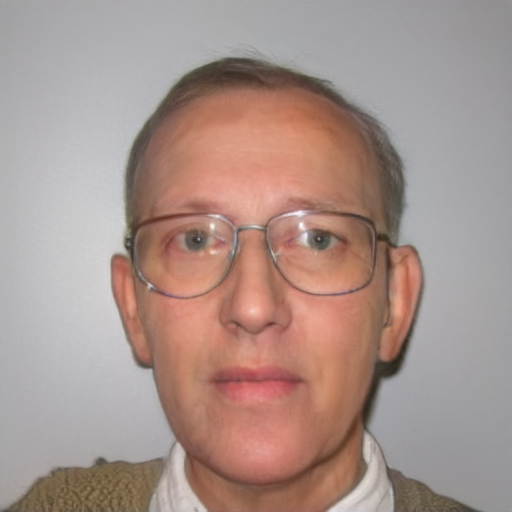}
\end{minipage}%
\begin{minipage}[b]{0.166\linewidth}
\includegraphics[width=\linewidth]{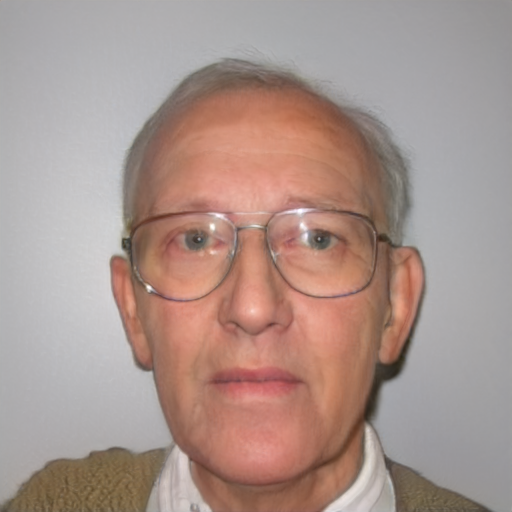}
\end{minipage}%
\begin{minipage}[b]{0.166\linewidth}
\includegraphics[width=\linewidth]{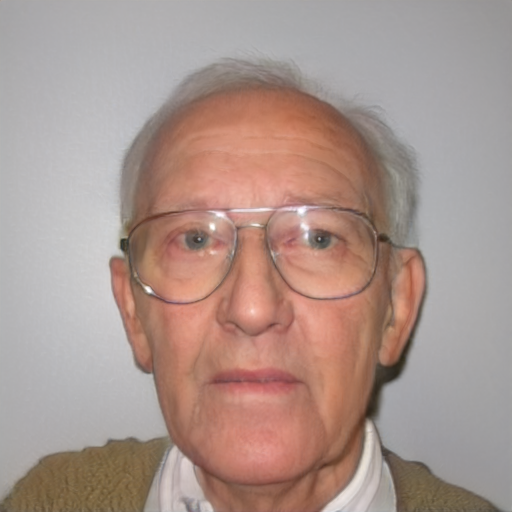}
\end{minipage}%
\begin{minipage}[b]{0.166\linewidth}
\includegraphics[width=\linewidth]{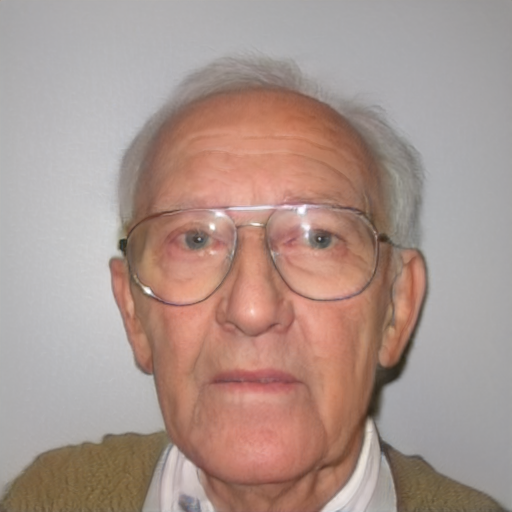}
\end{minipage}%
\begin{minipage}[b]{0.166\linewidth}
\includegraphics[width=\linewidth]{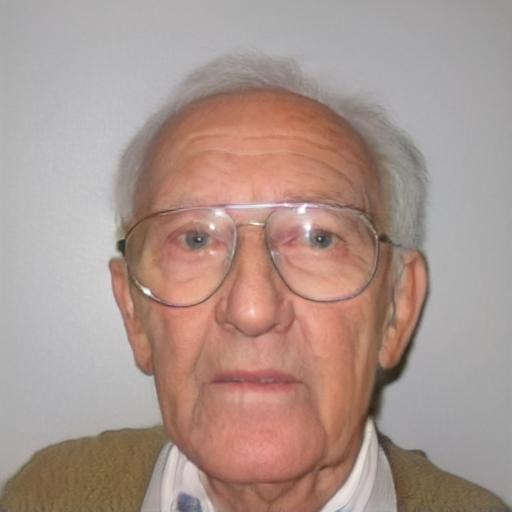}
\end{minipage}%
\begin{minipage}[b]{0.166\linewidth}
\includegraphics[width=\linewidth]{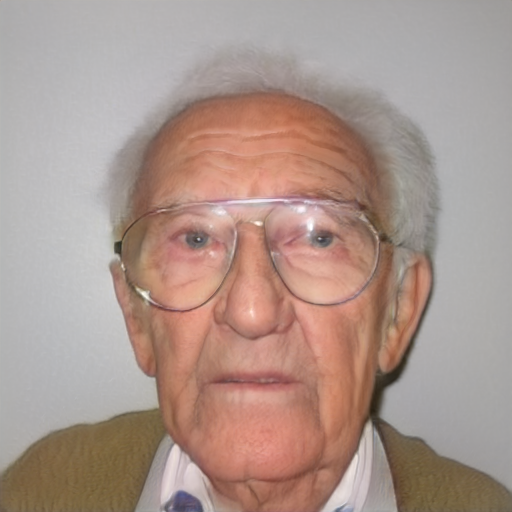}
\end{minipage}%


\caption{Exemplary patients' images generated by manipulating the age attribute.}
\label{Fig:IndividualImportanceFeature}
\end{figure}

 An observation that could not be neglected however is the similarity between progressing in age and worse prognostic features, as age is a common risk indicator. While one can assume an underlying correlation between age and general health related facial features, we addressed a potential bias of the model to only rely on age features for the prognosis prediction task. 
 As shown in Fig.\,\ref{Fig:JointFeatures}, it is clearly observable that when editing the patient's image based on the health attribute, distinct facial features are affected compared to when editing based on the age attribute, as shown in Fig.\,\ref{Fig:IndividualImportanceFeature}.

Based on the feature importance indicated by the regression model's coefficients $\vw$ in Eqn.(\ref{eqn:jointFeature}), changes in the embedding vector elements with corresponding higher coefficients have a more observable impact on the attributes of the generated images. This demonstrates that, as in Fig.\,\ref{Fig:JointFeatures}, the StyleGAN regression model is indeed using meaningful facial features instead of bias or random information for survival prediction.
 
 
\section{Conclusion}
This work investigates on prognosis prediction of cancer patients using portrait photos. Our work demonstrates that the latent features extracted by StyleGAN contain highly relevant prognostic information, and an effective regression model can be optimized for survival prediction. Such a model using facial features has good interpretability and can achieve comparable survival prediction accuracy to a state-of-the-art model using clinical record data. The constructed health attribute allows photo editing into a healthier appearance or prognosis of a more severe illness, which has important potential value for patient care. 

\subsubsection{\discintname}
The authors have no conflict of interest to declare for this work.
\bibliographystyle{splncs04}
\bibliography{ref}
\end{document}